# Stabilization of lead-free bulk CsSnI$_3$ perovskite thermoelectrics via incorporating of TiS$_3$ nanoribbon clusters

Alexandra Ivanova,[1,*] Lev Luchnikov,[1] Margarita Golikova,[1] Dmitry S. Muratov,[2] Danila Saranin,[1] Aleksandra Khanina,[1] Pavel Gostishchev,[1] Vladimir Khovaylo[1]

[1]National University of Science and Technology MISIS (NUST MISIS), Leninsky av. 4, Moscow, 119049, Russia
[2]Chemistry Department, University of Turin, 10125, Turin, Italy

**Abstract**

The intense research for efficient low-temperature thermoelectric materials motivates the exploration of innovative compounds and composite systems. This study examines the effects of integrating low-dimensional titanium trisulfide (TiS$_3$) into bulk tin-based halide perovskites (CsSnI$_3$) for use in thermoelectric applications. The addition of small amounts of two-dimensional titanium trisulfide (TiS$_3$) to bulk tin-based halide perovskites (CsSnI$_3$) significantly enhanced the structural stability of the composite material and suppressed oxidation processes. The CsSnI$_3$-TiS$_3$ composites demonstrated stabilization of temperature-dependent electrical properties (conductivity and Seebeck coefficient). This study provides valuable insights into the promising approach of using low-dimensional TiS$_3$ as an additive to stabilize the thermoelectric performance of CsSnI$_3$.

## Introduction

Research on low-temperature thermoelectric materials represents a significant stage in the development of modern thermoelectric conversion technologies.[1,2] Thermal and electrical energy may be used more efficiently for applications like radiation sensors, thermal electronic devices, thermoelectric cooling, and so on by increasing the figure of merit, $zT$, of a thermoelectric material which as defined as: $zT = \alpha^2 \sigma T/(\kappa_{lat} + \kappa_{el})$, where $\alpha$ represents the Seebeck coefficient, $\sigma$ denotes electrical conductivity, and $\kappa_{lat}$ and $\kappa_{el}$ represent lattice and electronic thermal conductivity, respectively.[3]

Halide perovskites represent a novel class of materials garnering attention in the fields of photovoltaics, optoelectronics, and thermoelectrics.[4–8] Inorganic tin-based perovskites have attracted interest due to simplified synthesis, high charge carrier mobility, and ultra-low thermal conductivity.[9] However, these materials also present challenges regarding stability, thereby opening avenues for further research and development.[10] Recently, we performed a complex analysis for the degradation dynamics of bulk CsSnI$_3$ perovskites. Bulk perovskites demonstrate enhanced stability under ambient conditions compared to thin-film samples. It is hypothesized that surface passivation of perovskites plays a key role in preserving *p*-type conductivity, even after prolonged exposure to air. The oxidation of bulk CsSnI$_3$ results in the formation of Cs$_2$SnI$_6$, which leads to a significant reduction in the power factor. Consequently, achieving a combined increase in the stability and efficiency of perovskite materials is critical challenge.

Significant research efforts have focused on stabilizing thin films of tin-based perovskites for use in opto-electronic and photoelectronic devices. It has been demonstrated that precise control over the synthesis and processing conditions of perovskites can significantly enhance material stability.[11,12] Partial substituting of the tin ions in the perovskite structure with germanium or lead, can enhance phase stability.[13,14] Additionally, encapsulating layers based on organic polymers or oxides have been applied to the surface of the perovskite layer, providing protection against exposure to moisture and oxygen.[15,16] The doping of tin-based perovskites was proposed as an effective strategy for relaxation of the strain in molecular structure.[17–19] The use of low-dimensional materials provides wide range of possibilities for surface passivation, interface modification and tuning of transport properties.[20–25]

Recently, metal tri-trisulfides attracted the attention of material scientists for application in electronics and optoelectronics.[26–34] Titanium trisulfide (TiS$_3$) has high potential in thermoelectric application due to its high Seebeck coefficient and lower thermal conductivity compared to transition metal dichalcogenides.[35–37] Rich surface properties of low dimensional materials provide large possibilities for interface modification as for thin-films materials as for bulk ones.

In this paper, we demonstrate a novel approach for stabilizing $CsSnI_3$ bulk thermoelectric materials through the formation of composites with $TiS_3$. To assess the impact of adding titanium trisulfide to tin-based perovskites, we prepared a series of bulk samples with specified nominal compositions of $CsSnI_3 + xTiS_3$ ($x$ = 0, 3, 5 and 7 wt.%). The maximum thermoelectric efficiency value observed in these composites was 0.055 for the sample with a composition of $CsSnI_3$ + 3 wt.% $TiS_3$. We observed that $CsSnI_3$ -$TiS_3$ composites exhibited consistent stabilization of temperature-dependent electrophysical properties. The potential mechanisms underlying stabilization in composite thermoelectrics and its impact on transport and recombination processes were discussed.

**Experimental**

**$CsSnI_3$.** The pure $CsSnI_3$ ingots were produced using the vacuum melting technique. CsI granules (with a purity of 99.998 %, from LLC Lanhit, Russia) and $SnI_2$ granules (with a purity of 99.999 %, from LLC Lanhit, Russia) were weighed and mixed in the stoichiometric ratios in an argon-filled (99.998 % purity) glove box to prevent any contamination or oxidation. These prepared mixtures were then securely sealed in evacuated quartz tubes with a 20 mm inner diameter and a 1.5 mm wall thickness. Subsequently, the mixture underwent controlled thermal treatment, gradually heating to 923 K at a rate of 100 Kh$^{-1}$, holding at this temperature for 24 hours, and then gradually cooling back to room temperature in the furnace. The opening of the quartz tubes post-melting was conducted in the argon-filled glove box.

**$TiS_3$.** The $TiS_3$ crystals were grown by the direct reaction between Ti shavings (99.99 % purity) and S powder (99.999 % purity). Raw materials were weighted in the stoichiometric ratios and sealed in separate part of an evacuated quartz tube, ~10 wt.% excess of S was added before melting. The tube was put in a tubular furnace in such a way that a temperature gradient could be created, with one end (with Ti) in the middle of the furnace, which is the hottest part, and the other end (with S) closer to the furnace opening. The furnace was heated up to 873 K at the rate of 300 Kh$^{-1}$ and the material was annealed for 5 days. The furnace was shut off when the synthesis was finished, and as the ampule cooled, the excess S gathered at the cooler end.

**$CsSnI_3 + xTiS_3$ ($x$ = 0, 3, 5 and 7 wt.%).** The composites were mixed in an agate mortar in the argon-filled glove box. The powders were cold pressed under a uniaxial stress of 250 MPa for 5 minutes in a cylindrical high strength stainless steel die with an internal diameter of 10 mm. Then, the compacted samples were sealed in an evacuated quartz tubes and pressureless sintered at 433 K for 5 hours.

**Characterization.** X-ray diffraction (XRD) patterns were obtained by using a TDM-20 diffractometer (Dandong Tongda Science & Technology Co., Ltd., China) with Cu-K$\alpha$ radiation ($\lambda$ = 1.5419 Å). To further examine the morphology and chemical composition of the bulk specimens scanning electron microscopy (SEM; Vega 3 SB, Tescan, Czech Republic) and energy dispersive X-ray spectroscopy (EDX; x-act, Oxford Instruments, UK) were employed. Additionally, Raman scattering data were acquired using a Thermo DXR spectrometer, with excitation provided by a 532 nm laser source.

**Transport property measurements.** Sample preparation for measurements of electrical transport properties was carried out in an argon-filled glovebox. Consolidated pellets were cut into bars (3x10x2 mm$^3$) using a hand-held string saw. Samples were transferred to the setup without exposure to air in the glovebox. The evaluation included the simultaneous determination of both electrical conductivity $\sigma$ and the Seebeck coefficient $\alpha$, employing four-probe and differential methods, respectively. The measurements were performed under a helium atmosphere in a temperature range from 280 to 420 K. These analyses were carried out within laboratory-made systems situated at the National University of Science and Technology MISIS. The total thermal conductivity, $\kappa$, was calculated from the measured thermal diffusivity, $\chi$, specific heat, $C_p$, and density $d$ using the well-known relationship $\kappa = \chi C_p d$. The density of the samples was measured by the Archimedes principle in isopropyl alcohol, which does not dissolve perovskite. The specific heat of the matrix $CsSnI_3$, $(C_p)_m$, and that of the $TiS_3$ inclusions, $(C_p)_i$, was calculated in the framework of the Debye model. $C_p$ of the $CsSnI_3 + xTiS_3$ ($x = 0$, 3, 5 and 7 wt.%) composites were calculated using the rule of mixtures, which is based on the assumption that a composite material behaves as a homogeneous material with properties that are a weighted average of its components. In this approach the specific heat of the composite material is calculated as $C_p = f(C_p)_i + (1-f)(C_p)_m$, where $f$ is the mass fraction of each component.[38] The thermal diffusivity was measured by the laser flash method using a LFA 447 NanoFlash (Netzsch, Germany). The combined uncertainty for all measurements involved in the $zT$ calculation is 16 %.

### Results and discussion

Powder X-ray diffraction analysis for the samples was done during the synthesis stage (**Fig.S1**). Phase analysis revealed that the prepared powders correspond to black orthorhombic perovskite $CsSnI_3$ (PDF# 01-080-2139) (**Fig.S1(a)** in **S.I.**) and titanium trisulfide corresponds to its monoclinic phase (PDF# 00-036-1337) with space group P21/m (**Fig.S1(b)** in **S.I.**).

For all samples with compositions $CsSnI_3 + x$ wt.% $TiS_3$ ($x = 0$, 3, 5 and 7 wt.%) their phase composition was investigated depending on the exposure time in air (**Fig. 1**). The exposure conditions in air were within 17–21 ºC, 27–35 % relative humidity. The main reflections in all samples corresponded to the orthorhombic black phase of $CsSnI_3$ (PDF# 01-080-2139). Additionally, the yellow phase Y-$CsSnI_3$ (PDF# 01-071-1898) was identified on the surfaces of the samples. The presence of the yellow perovskite modification is explained by the fact that sample mounting and diffraction pattern acquisition were conducted in ambient air. Initial observation revealed that the quantity of yellow perovskite modification is independent of the amount of added titanium trisulfide (**Table 1**). In the sample with 3 wt.% $TiS_3$, the minimal surface content of the yellow modification was 5 %, whereas in the additive-free perovskite, it constituted 20 %. Notably, after one hour of air exposure, the yellow modification Y-$CsSnI_3$ in the additive-free perovskite increased by 10 % to a total of 30 % of the surface phases (**Fig. 1(e)**). In the sample with 3 wt.% $TiS_3$, the yellow phase Y-$CsSnI_3$ increased by 2 %, while in samples with 5 wt.% and 7 wt.%, the quantity of yellow phase Y-$CsSnI_3$ remained unchanged relative to the non-exposed composition. Examination after 24 hours of air exposure revealed differing degradation mechanisms among the samples (**Fig. 1(f)**). After 24 hours of air exposure, the additive-free sample predominantly exhibited black orthorhombic modification $CsSnI_3$, constituting 58 %. Double perovskite $Cs_2SnI_6$ accounted for 24 % on the surface, with only 18 % yellow modification Y-$CsSnI_3$. A similar trend was observed in the 5 wt.% $TiS_3$ sample. After 24 hours of air exposure, the surface of the bulk sample retained 23 % black orthorhombic modification. A characteristic feature of these samples is the initial presence of a yellow modification on the surface, ranging from 15-20 %. In bulk perovskites obtained through consolidation by pressureless sintering, this significant quantity of surface yellow modification likely facilitated the rapid formation of double perovskite. The surface modification formed stabilized buffer interlayer, delaying further transitions within the bulk material.

The incorporation of titanium trisulfide into the perovskite structure effectively suppressed structural degradation for 24 hours. The $CsSnI_3$-$TiS_3$ composites exhibited significantly reduced dynamics of double perovskite formation, which, in contrast to the yellow modification, does not revert to the black orthorhombic phase.

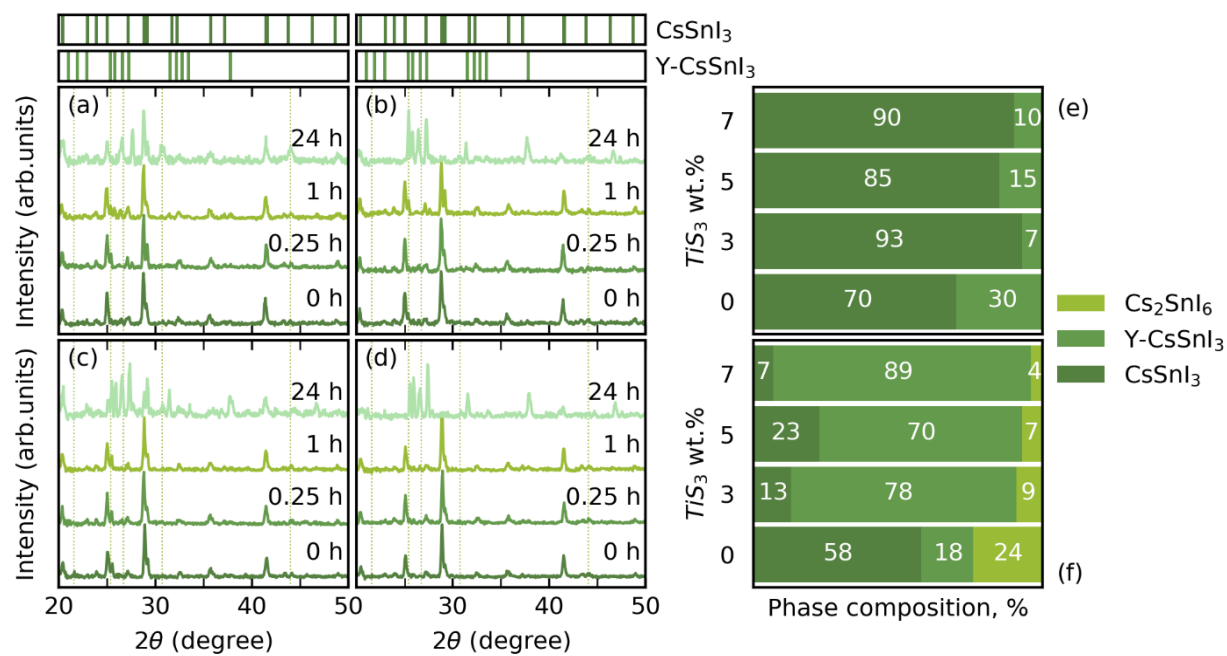

**Figure 1.** XRD patterns of the samples with a composition of (a) $CsSnI_3$, (b) $CsSnI_3$ + 3 wt.% $TiS_3$, (c) $CsSnI_3$ + 5 wt.% $TiS_3$, (d) $CsSnI_3$ + 7 wt.% $TiS_3$ following exposure intervals in ambient air. The dashed lines correspond to the $Cs_2SnI_6$. The phase composition of the $CsSnI_3$ + $x$ wt.% $TiS_3$ ($x$ = 0, 3, 5 and 7 wt.%) samples following (e) 1 hour and (f) 24 hours in ambient air

To investigate the phase composition on the surface of $CsSnI_3$-$TiS_3$ composite we performed Raman spectroscopy measurements. Raman spectra of pristine $TiS_3$ crystals, pure $CsSnI_3$ tablet and $CsSnI_3$ + $TiS_3$ composite are shown in **Fig. 2**. Two different spectra of composites were measured close and far from $TiS_3$ clusters. Relative intensities of 78 and 125 cm$^{-1}$ bands, which were attributed to $Cs_2SnI_6$, are higher in the areas farther from $TiS_3$ phase, suggesting that a formation of $Cs_2SnI_6$ close to $TiS_3$ is significantly impaired. Most likely this could be due to the evolution of $H_2S$ gas from $TiS_3$ particles in presence of ambient water. After 24 hours of exposure to air, the surface of the unmodified $CsSnI_3$ tablet exhibited regions enriched with a yellow phase. The presence of $TiS_3$ improved the phase composition uniformity on surfaces of samples containing 3-7 wt.% $TiS_3$, as shown in **Fig.S2** in **S.I.** Raman spectral analysis of $CsSnI_3$ in **Fig.S3** in **S.I.** revealed the presence of peaks at 105 and 150 cm$^{-1}$, which are characteristic of the yellow phase sites and may be associated with $SnI_4$ or $Sn(OH)_4$.[39] The $CsSnI_3$-$TiS_3$ composites showed no signals suggesting decomposition or oxidation products.

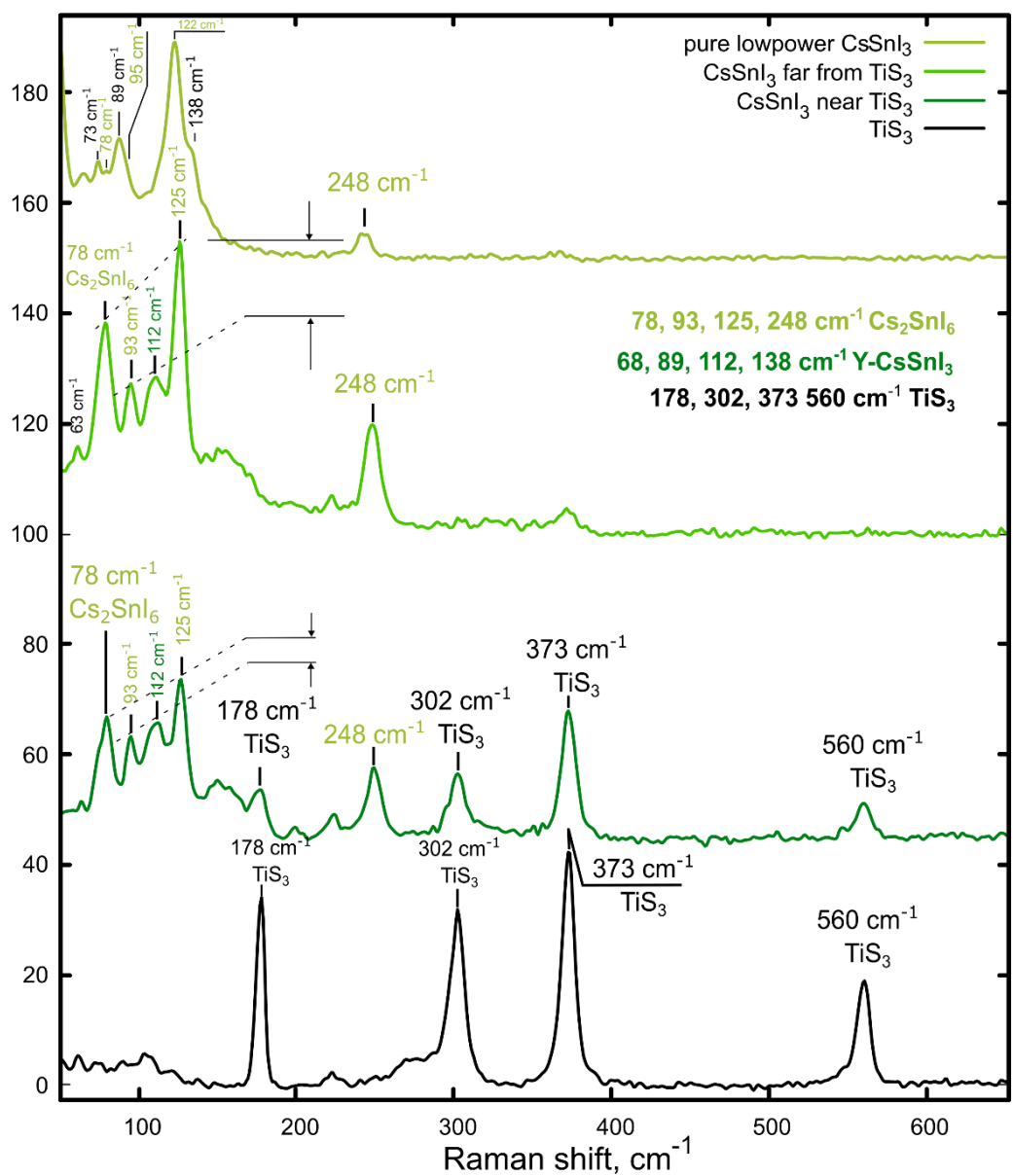

**Figure 2.** Raman spectra of pure CsSnI$_3$, areas far and near to TiS$_3$ clusters and pristine TiS$_3$ powder, shown from top to bottom respectively

The relative density of all specimens after sintering exceeded 95 % of the theoretical density (**Table 1**). The theoretical density of the composites was calculated using the same method as the specific heat capacity, employing the rule of mixtures.[38]

SEM imaging and EDX surface mapping of sintered samples without exposure to air indicate distribution of TiS$_3$ particles across the samples, exhibiting a broad range of sizes (**Fig.S4-S7** in **S.I.**). In the microstructure of the perovskite, both fine (~10 µm) and larger (up to ~400 µm) clusters of TiS$_3$ are observed. EDX analysis results demonstrate a good correlation between the actual composition of the samples and the nominal composition (**Table 1**). Upon normalizing the values to one cesium and one titanium in the actual compositions, it was found that all samples exhibit an excess of iodine and a deficit of sulfur. Furthermore, EDX analysis of the matrix indicates the presence of sulfur in areas adjacent to the TiS$_3$ inclusions. Nevertheless, no elemental substitutions were detected via EDX analysis. Notably, samples containing TiS$_3$ emit a distinctive odor of hydrogen sulfide, suggesting the potential presence of sulfur in the form of hydrogen sulfide on their surfaces.

**Table 1.** Nominal and actual composition (from the EDX analysis), phase composition and relative density $d$ of the CsSnI$_3$ + $x$ wt.% TiS$_3$ ($x = 0$, 3, 5 and 7 wt.%) without exposure in air.

| Nominal composition | Actual composition | Phase composition (vol %) | d, g/cm$^3$ (%) |
|---|---|---|---|
| CsSnI$_3$ | Cs$_1$Sn$_{1.1}$I$_{3.21}$ | 80 % CsSnI$_3$, 20 % Y-CsSnI$_3$ | 4.5 (99.6) |
| CsSnI$_3$ + 3 wt.% TiS$_3$ | Cs$_1$Sn$_{1.11}$I$_{3.24}$ + Ti$_1$S$_{2.87}$ | 95 % CsSnI$_3$, 5 % Y-CsSnI$_3$ | 4.43 (98.8) |
| CsSnI$_3$ + 5 wt.% TiS$_3$ | Cs$_1$Sn$_{1.07}$I$_{3.29}$ + Ti$_1$S$_{2.91}$ | 85 % CsSnI$_3$, 15 % Y-CsSnI$_3$ | 4.28 (95.9) |
| CsSnI$_3$ + 7 wt.% TiS$_3$ | Cs$_1$Sn$_{1.14}$I$_{3.34}$ + Ti$_1$S$_{2.86}$ | 90 % CsSnI$_3$, 10 % Y-CsSnI$_3$ | 4.24 (95.6) |

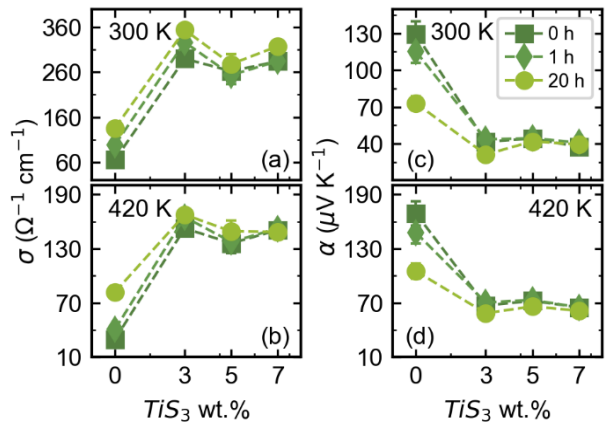

**Figure 3.** Electrical conductivity at (a) 300 K and (b) 420 K and Seebeck coefficient at (c) 300 K and (d) 420 K as a function of TiS$_3$ content

The electrical transport properties of the composites were measured over several heating and cooling cycles to confirm the reproducibility of the data within the range of 300–420 K (**Fig.S8 in S.I.**). To assess material stability, measurements were conducted on samples both without exposure and with exposure to air for 1 hour and 20 hours. It can be observed that the stability/reproducibility of the data increases proportionally with the increase in the amount of TiS$_3$ inclusions in the composites. This is more clearly demonstrated by concentration dependencies (**Fig.3**). In the sample without TiS$_3$ addition, the electrical conductivity at room temperature increases by 50 % after 1 hour and by 100 % after 20 hours of exposure to air (**Fig.3(a)**). This is attributed to the self-doping process in the tin-based CsSnI$_3$ perovskites, which leads to an increase in charge carrier concentration and, consequently, in electrical conductivity.[40] This process also explains the inverse behavior of the Seebeck coefficient, i.e., the decrease in its absolute values with increasing sample exposure to air (**Fig.3(c)**). The value of the Seebeck coefficient at room temperature in the sample without TiS$_3$ decreased by 11 % after 1 hour in air and by 43 % after 20 hours. With the addition of 3 wt.% TiS$_3$, the change in conductivity values after 1 hour decreased to 13 % and to 22 % after 20 hours of exposure to air (**Fig.3(a)**, **Fig. S6(c)** in **S.I.**). The Seebeck coefficient after 1 hour in air is within the measurement error, and after 20 hours, it decreases by 26 %. Already with the addition of 5 wt.% TiS$_3$ to CsSnI$_3$ perovskite, stable values of both electrical conductivity and the Seebeck coefficient are achieved even after 20 hours exposure in air (**Fig. 3**, **Fig.S6(e, f)** in **S.I.**). The obtained values of electrical properties vary within the measurement error range across the entire temperature interval (~ 7 %). In the samples with 7 wt.% TiS$_3$ addition, at room temperature, the difference between the values without exposure and after exposure for 20 hours reaches 10 %. However, as the temperature rises, this difference decreases to 3–5 % (**Fig. 3(b, d)**, **Fig.S6(g, h)** in **S.I.**). Thus, despite the uneven distribution of trisulfide in the samples, their electrical properties remain stable within 20 hours of exposure to air.

However, the addition of TiS$_3$ also leads to a significant change in the electrical transport properties compared to the perovskite without additives. TiS$_3$ is an *n*-type semiconductor ($α$ = -650 $μ$VK$^{-1}$ at 300 K),[41] thus its addition reduces the Seebeck coefficient by 3 times (from 129 $μ$VK$^{-1}$ to ~ 40 $μ$VK$^{-1}$ at 300 K) compared to the pristine perovskite due to charge carrier recombination processes. This recombination process leads to a reduction in the chemical potential disparity between regions with differing carrier concentrations. Despite a fourfold augmentation in electrical conductivity (from 65 $Ω^{-1}$cm$^{-1}$ to ~ 280 $Ω^{-1}$cm$^{-1}$ at 300 K), this enhancement fails to fully offset the decline in the Seebeck coefficient. Consequently, the power factor, $α^2σ$, of the composites at room temperature is halved in comparison to the sample lacking additives (**Fig.4a**).

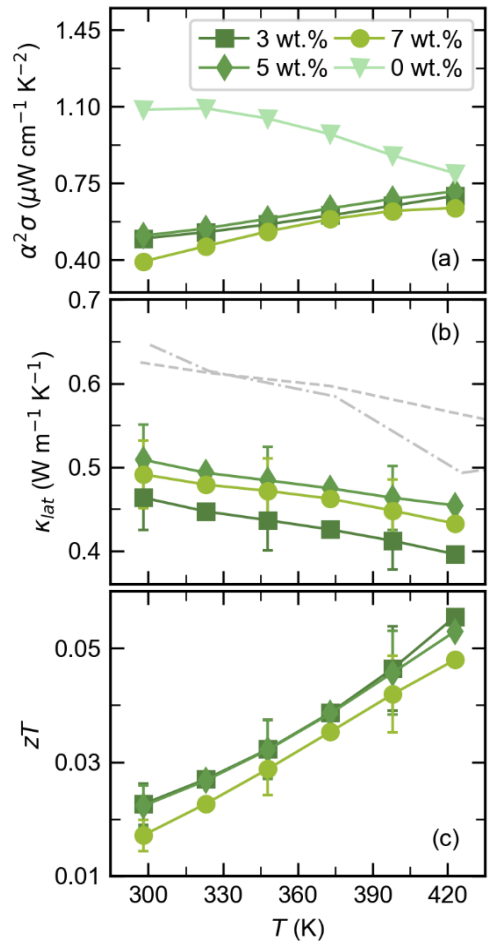

**Figure 4**. Temperature dependence of (a) the power factor $α^2σ$, (b) the lattice thermal conductivity $κ_{lat}$, and the thermoelectric efficiency $zT$ for the CsSnI$_3$ + *x* wt.% TiS$_3$ ($x$ = 0, 3, 5 and 7 wt.%). Literature data of lattice thermal conductivity for other CsSnI$_3$ perovskites are also shown for comparison (Qian et al.,[14] Yu et al.[42]).

Incorporation of TiS$_3$ leads to a reduction in the lattice component of thermal conductivity (**Fig. 4(b)**). All values range within 0.46–0.51 Wm$^{-1}$K$^{-1}$ at 300 K, which is approximately 26 % lower than that of the perovskite without TiS$_3$ (0.6–0.65 Wm$^{-1}$K$^{-1}$ at 300 K).[14,42] The minimum value at 300 K, 0.46 Wm$^{-1}$K$^{-1}$, is exhibited by the sample CsSnI$_3$ + 3 wt.% TiS$_3$. Increasing the amount of TiS$_3$ in the composite results in an increase in the lattice thermal conductivity, which may be associated with the higher thermal conductivity of the TiS$_3$ itself (~ 10 Wm$^{-1}$K$^{-1}$ at 300 K)[30,43] and the sizes of its clusters within the sample. The temperature dependence of the total thermal conductivity of the samples exhibits a similar trend to the lattice thermal conductivity (**Fig.S9(a)**). As the addition of TiS$_3$ within the range of 3 to 7 wt.% in the composites does not significantly affect the electrical transport properties (**Fig.3**), the electrical thermal conductivity for all samples is within the measurement error (**Fig.S9(b)**). Additionally, the contribution of the lattice component predominates in the total thermal conductivity for all samples and constitutes 70-80 %.

The thermoelectric figure of merit $zT$ of all samples increases with temperature (**Fig. 4(c)**). Throughout the measured temperature range, the values of the entire series $CsSnI_3 + x$ wt.% $TiS_3$ ($x$ = 3, 5, and 7 wt.%) are within the measurement error. The maximum efficiency value reached $zT$ = 0.055 at 420 K for the sample with the composition $CsSnI_3$ + 3 wt.% $TiS_3$.

**Conclusions**

In summary, new composites of $CsSnI_3 + x$ wt.% $TiS_3$ ($x$ = 0, 3, 5, and 7 wt.%) were synthesized by vacuum melting and chemical vapor transmission process following by pressureless sintering. Incorporating titanium trisulfide ($TiS_3$) into bulk tin-based halide perovskites ($CsSnI_3$) for thermoelectric applications induces significant changes in electrical transport properties. Although $TiS_3$ improves electrical conductivity, it reduces the Seebeck coefficient by facilitating charge carrier recombination processes, consequently halving the power factor compared to pristine perovskite samples. The addition of $TiS_3$ reduces the lattice component of thermal conductivity by approximately 26 %, contributing to improved overall thermal performance. When the $TiS_3$ content goes beyond 3 wt.%, the lattice thermal conductivity increases as a result of both the higher thermal conductivity of $TiS_3$ itself and the sizes of its clusters. Additionally, the presence of $TiS_3$ results in a 26 % reduction in the lattice component of thermal conductivity, significantly enhancing thermal performance. Increasing the $TiS_3$ content beyond 3 wt.% leads to higher lattice thermal conductivity, likely due to the inherently greater thermal conductivity of TiS3 and the sizes of its clusters in the sample. The maximum efficiency value of $zT$ = 0.055 at 420 K is achieved by the composite with 3 wt.% $TiS_3$. The incorporation of $TiS_3$ positively impacts material stability, as demonstrated by the consistent temperature-dependent electrophysical properties maintained over a 20-hour period. This enhanced stability is essential for practical applications, particularly in environments where thermoelectric devices undergo prolonged operational periods. The investigation underscores the potential of integrating low-dimensional materials like $TiS_3$ into bulk perovskites for tailored thermoelectric applications. Future research may explore optimization strategies to further enhance the performance of these composite systems, paving the way for next-generation thermoelectric conversion technologies.

**Author contribution**

**Alexandra Ivanova:** Methodology, Formal analysis, Investigation, Data curation, Visualization, Writing (Original draft), Writing (Review & Editing). **Lev Luchnikov:** Formal analysis, Investigation, Data curation, Writing (Original draft), Writing (Review & Editing). **Margarita Golikova:** Investigation, Data curation. **Dmitry S. Muratov:** Conceptualization, Methodology, Formal analysis, Visualization, Writing (Original draft), Writing (Review & Editing). **Danila Saranin:** Resources, Supervision, Writing (Review & Editing). **Aleksandra Khanina:** Investigation. **Pavel Gostishchev:** Project administration, Funding acquisition. **Vladimir Khovaylo:** Supervision, Writing (Review & Editing).

**Conflicts of interest**

There are no conflicts to declare.

**Data Availability Statement**

The data that supports the findings of this study are available within the article and its supplementary material.

**Acknowledgments**

The study was carried out with financial support from the Russian Science Foundation (project no. 22-79-10326).

# Stabilization of lead-free bulk $CsSnI_3$ perovskite thermoelectrics via incorporating $TiS_3$ nanoribbon clusters

Alexandra Ivanova,[1,*] Lev Luchnikov,[1] Margarita Golikova,[1] Dmitry S. Muratov,[2] Danila Saranin,[1] Aleksandra Khanina,[1] Pavel Gostishchev[1] and Vladimir Khovaylo[1]

[1] National University of Science and Technology MISIS (NUST MISIS), Leninsky av. 4, Moscow, 119049, Russia

[2] Chemistry Department, University of Turin, 10125, Turin, Italy

[*] Corresponding author. E-mail address: aivanova@misis.ru (A. Ivanova).

**XRD data**

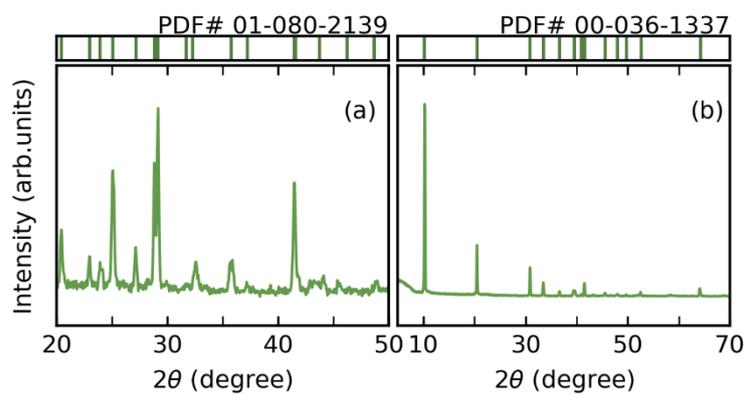

**Fig.S1.** PXRD patterns for (a) CsSnI$_3$ and (b) TiS$_3$ after synthesis.

**Raman data**

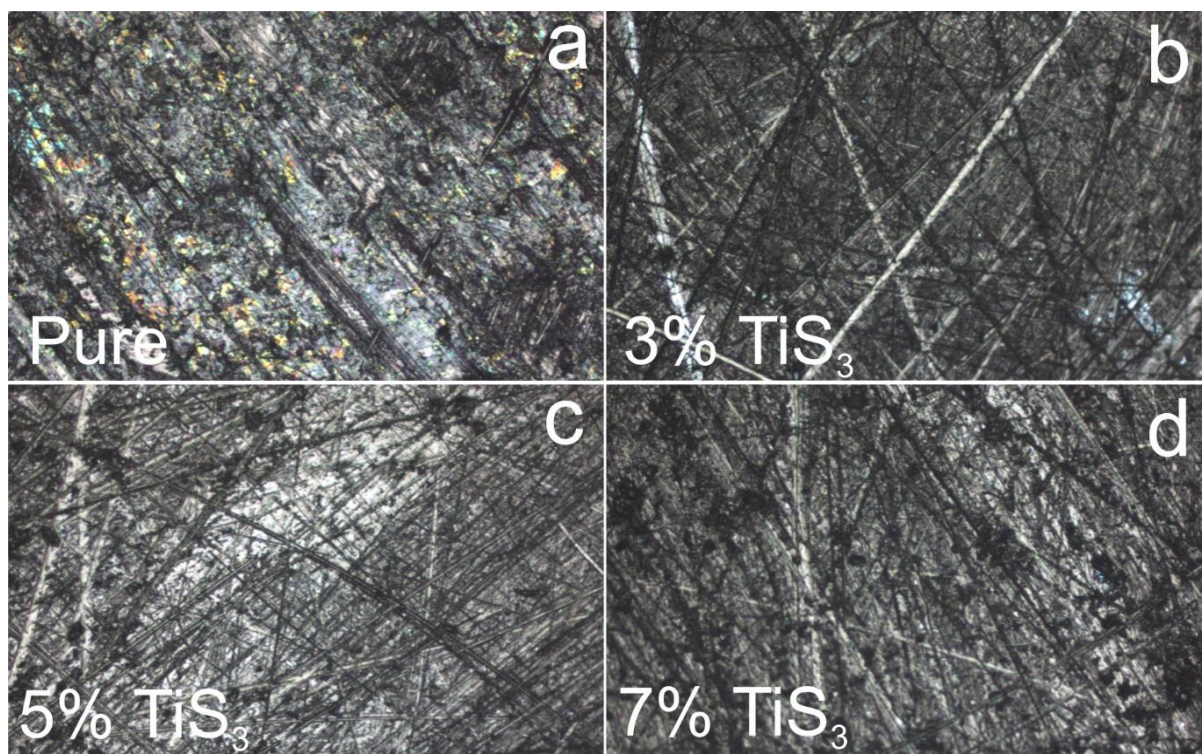

**Fig.S2.** Microphotography of CsSnI$_3$ tablets surfaces after 24h air exposure.

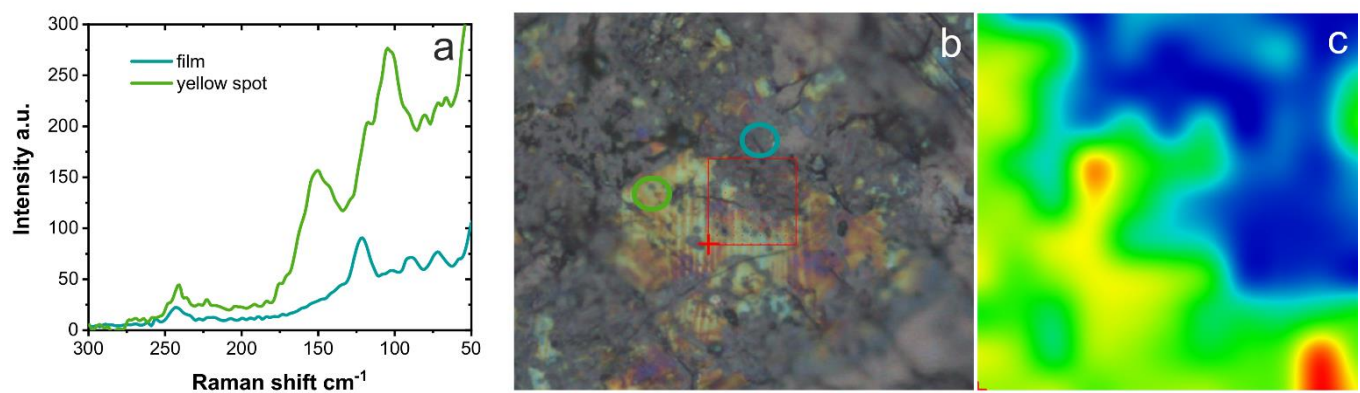

**Fig.S3.** Raman spectra (a) and raman map (b,c) of yellow-black region on surface of pure CsSnI$_3$ tablet after 24 air exposure.

**SEM images and EDX mapping**

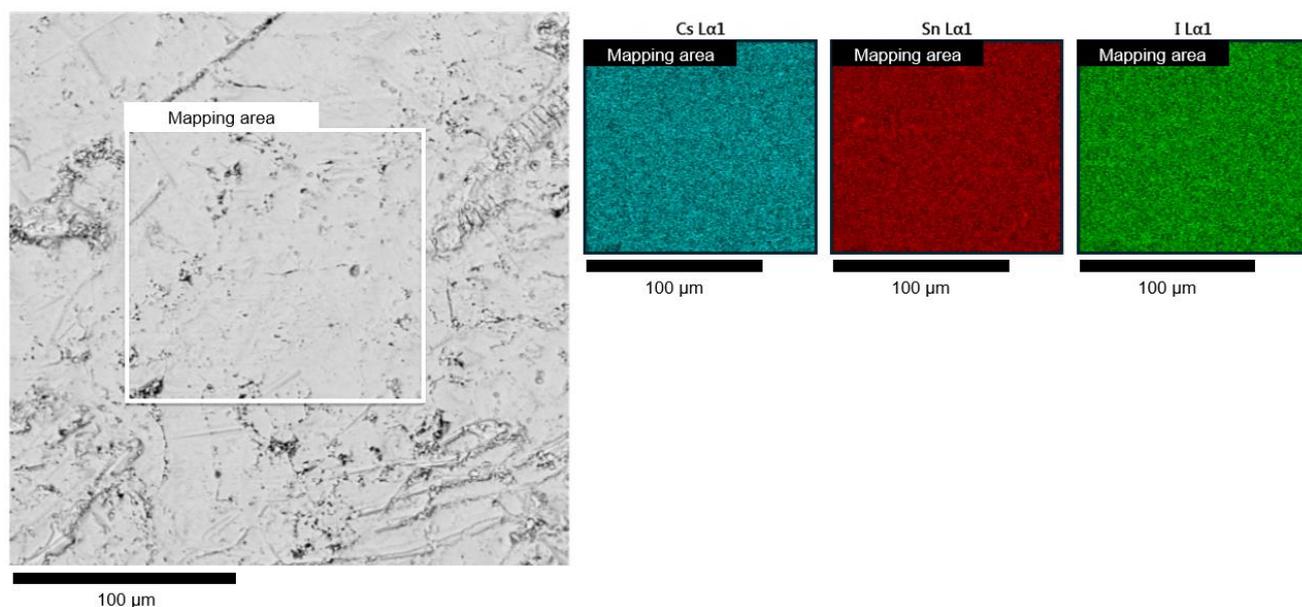

**Fig. S4.** SEM image of the polished surface of the CsSnI$_3$ without air exposure specimen and corresponding EDX maps.

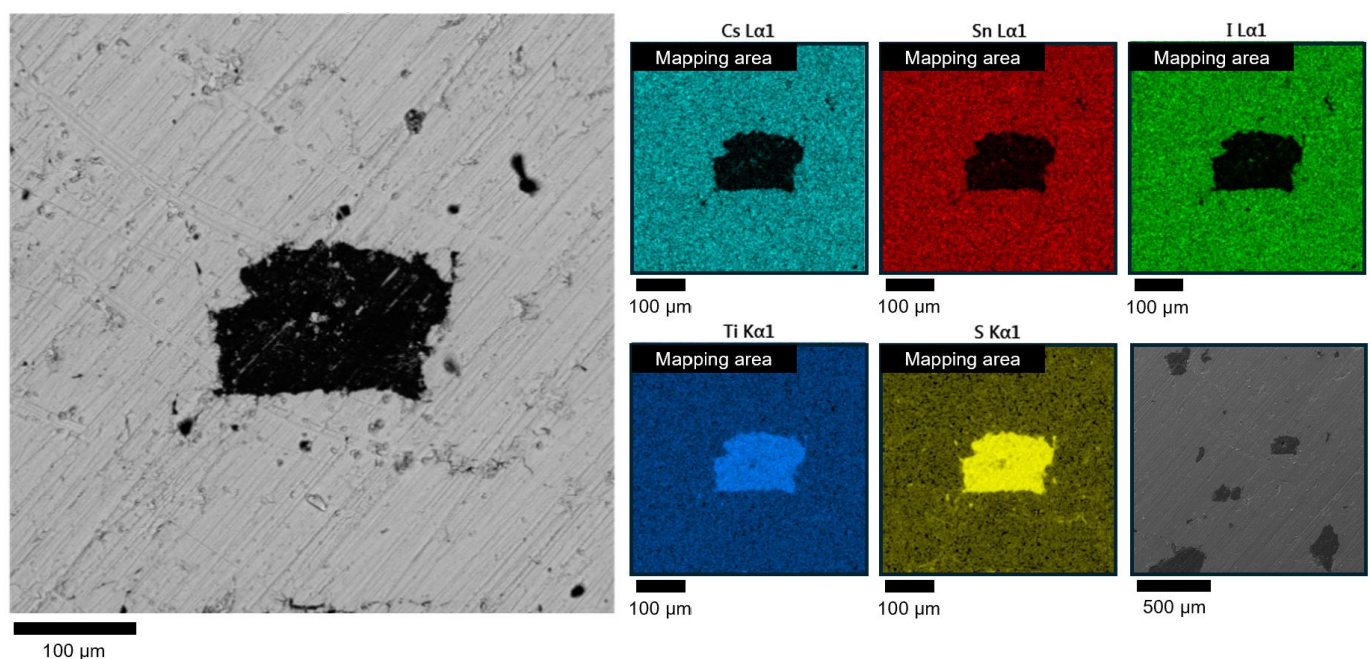

**Fig. S5.** SEM image of the polished surface of the CsSnI$_3$ + 3 wt. % TiS$_3$ without air exposure specimen and corresponding EDX maps.

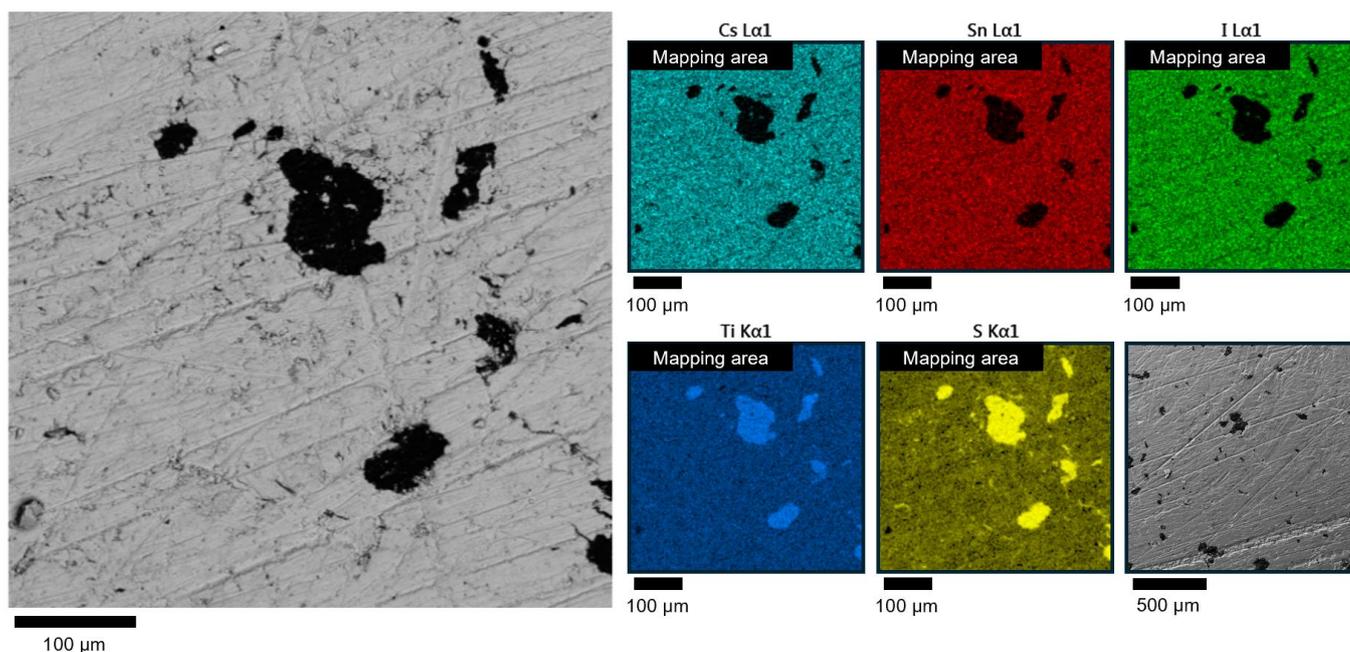

**Fig. S6.** SEM image of the polished surface of the CsSnI$_3$ + 5 wt. % TiS$_3$ without air exposure specimen and corresponding EDX maps.

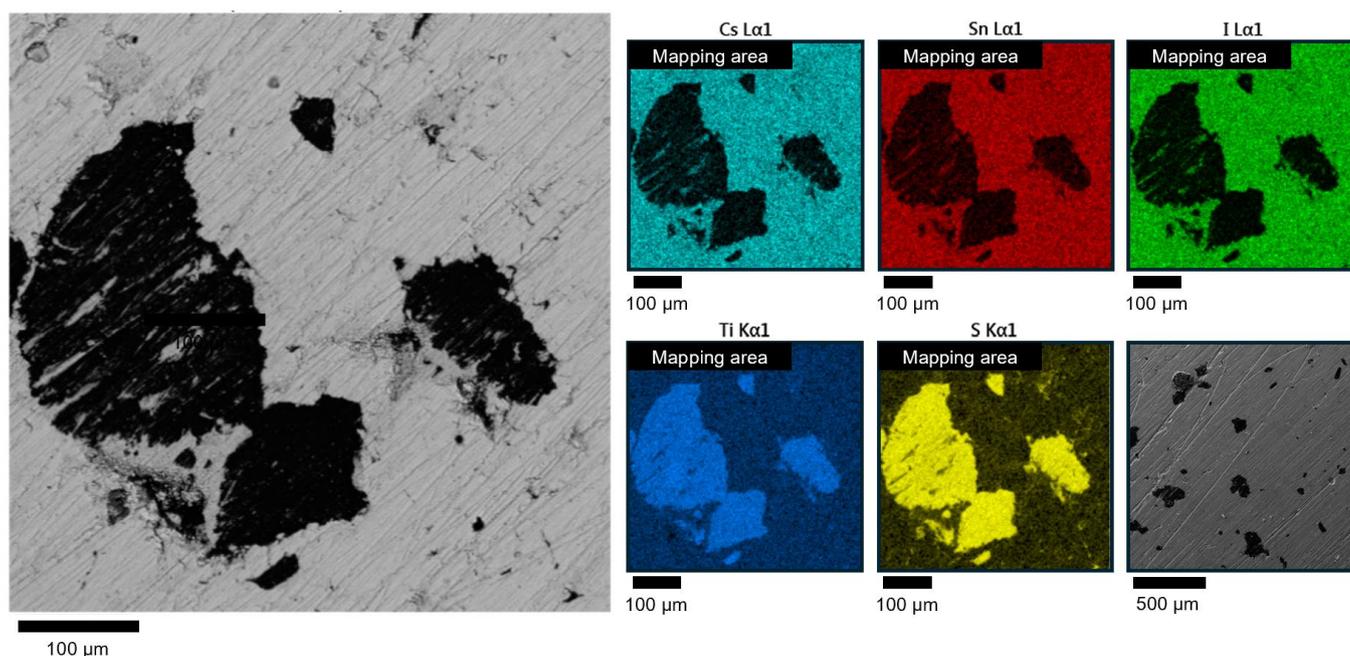

**Fig. S7.** SEM image of the polished surface of the CsSnI$_3$ + 7 wt. % TiS$_3$ without air exposure specimen and corresponding EDX maps.

**Electrical transport properties**

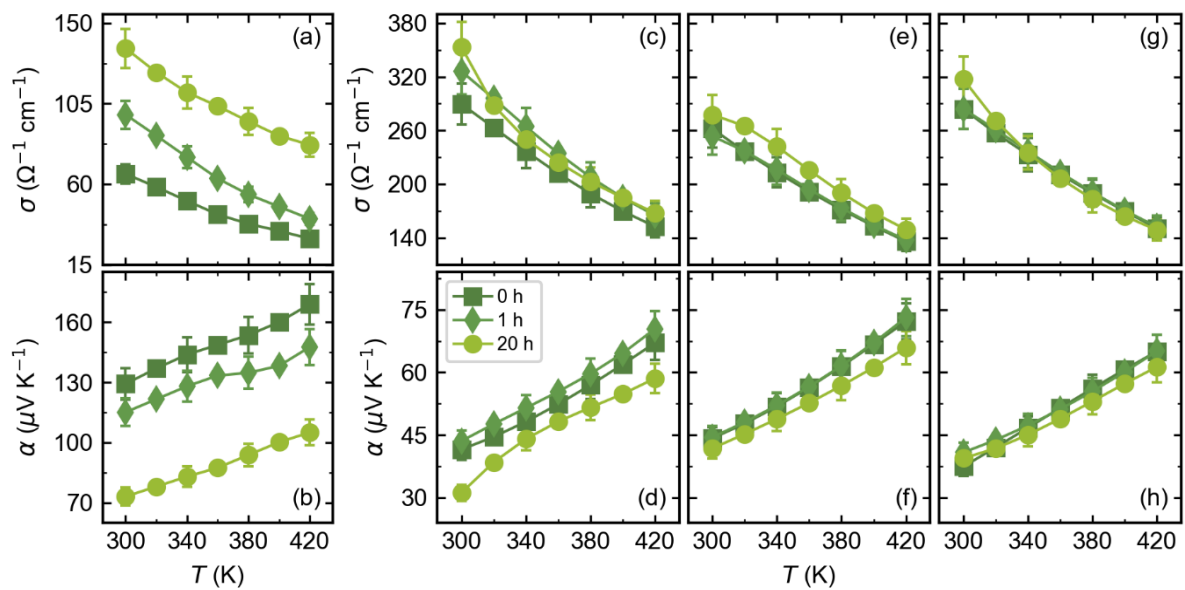

**Fig. S8.** Temperature dependence of electrical conductivity $\sigma$ and Seebeck coefficient $\alpha$ of (a, b) $CsSnI_3$, (c, d) $CsSnI_3$ + 3 wt.% $TiS_3$, (e, f) $CsSnI_3$ + 5 wt.% $TiS_3$, (g, h) $CsSnI_3$ + 7 wt.% $TiS_3$.

**Thermal transport properties**

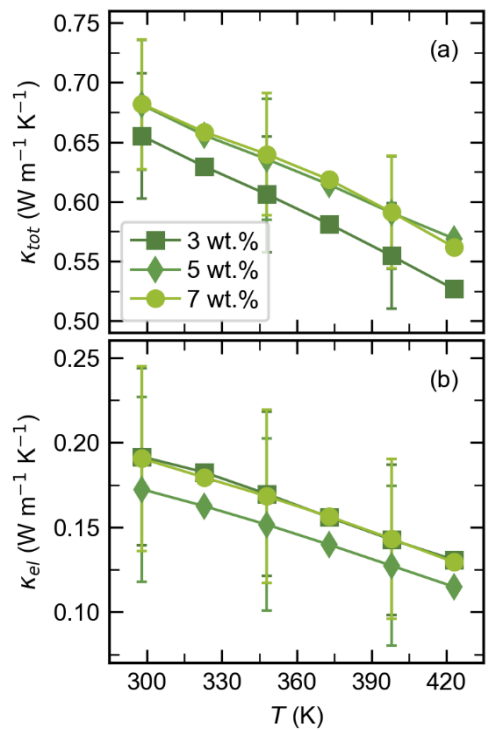

**Fig. S9.** Temperature dependence of (a) the total thermal conductivity and (b) the electrical thermal conductivity of the CsSnI$_3$ + $x$ wt.% TiS$_3$ ($x$ = 3, 5 and 7 wt.%)